\documentclass[
 reprint,
 amsmath,amssymb,
 aps, superscriptaddress, pre
]{revtex4-1}
\usepackage{graphicx}
\usepackage{amsmath}
\usepackage{amssymb}
\usepackage{enumerate}
\usepackage{bm}
\usepackage[justification=raggedright,singlelinecheck=false]{caption}
\usepackage[justification=raggedright,singlelinecheck=false]{subcaption}
\usepackage{cleveref}
\usepackage{verbatim}
\usepackage{makecell}
\usepackage{xcolor}
\usepackage{tikz}
\definecolor{darkGreen}{rgb}{0,0.6,0}

\def \be{\begin{equation}}
\def \ee{\end{equation}}
\def \ba{\begin{array}}
\def \ea{\end{array}}
\def \bea{\begin{eqnarray}}
\def \eea{\end{eqnarray}}

\begin{document}


\title{Normal form for renormalization groups}

\author{Archishman Raju}
\author{Colin B. Clement}
\author{Lorien X. Hayden}
\author{Jaron P. Kent-Dobias}
\author{Danilo B. Liarte}
\author{D. Zeb Rocklin}
\author{James P. Sethna}
\email{sethna@lassp.cornell.edu}

\affiliation{Laboratory of Atomic and Solid State Physics, Cornell University, Ithaca, NY 14853-2501, USA}

\begin{abstract}
The results of the renormalization group are commonly advertised as the existence of power law singularities near critical points. The classic predictions are often violated and logarithmic and exponential corrections are treated on a case-by-case basis. We use the mathematics of normal form theory to systematically group these into \textit{universality families} of seemingly unrelated systems united by common scaling variables. We recover and explain the existing literature and predict the nonlinear generalization for the universal homogeneous scaling functions. We show that this procedure leads to a better handling of the singularity even in classic cases and elaborate our framework using several examples. \end{abstract}
 \date{\today}


\maketitle
\section{Introduction}

Emergent scale invariance is a key to many of our current scientific and engineering challenges, including cell membranes~\cite{machta2012critical}, turbulence~\cite{canet2016fully}, fracture and plasticity~\cite{shekhawat2013damage, chen2013scaling}, and also the more traditional continuous thermodynamic phase transitions. 
The current formulation of the field has an elegant framework which can explain observables that scale as power laws times homogeneous functions. However, the literature on corrections to this result, including logarithms and exponentially diverging quantities, is much more scattered and does not have a similarly systematic framework. 

The renormalization group (RG) is our tool for understanding emergent
scale invariance. At root, despite challenges of implementation, the renormalization group (RG)
coarse grains and rescales the system to generate ordinary differential equations (ODEs) for model parameters as a
function of the observed log length scale $\ell$. A fixed point of these
flows represents a system which looks the same at different length scales;
systems near criticality flow near to this fixed point. In cases where the
flow can be linearized around the fixed point, the RG
implies that observables near criticality are given by a power law times
a universal function of an invariant combinations of variables; {\em e.g.}
the Ising model has magnetization $m \sim t^{\beta} \mathcal{M}(L t^\nu)$ where $L$ is the system size and $t = (T - T_c)/T_c$ is the deviation of the temperature $T$ from the critical temperature $T_c$.

Surprisingly often, this scenario of universal critical exponents
and scaling functions is violated; free energies and correlation lengths
scale with logarithms or exponentials, and the proper form of the 
universal scaling functions is often unknown. 
Specifically,
deviations arise in the Ising model in $d = 1$~\cite{ising1925beitrag}, 2~\cite{salas2002exact}, \& 4~\cite{larkin1995phase},
the tricritical Ising model in $d = 3$~\cite{Wegner73}, the $d=2$ XY
model~\cite{KosterlitzT73}, the surface critical behavior of
polymers~\cite{Diehl87, Eisenriegler88}, van der Waals interactions in 3-d spherical model~\cite{Dantchev06}, finite size scaling of the random field
Ising model (RFIM) in $d = 6$~\cite{Ahrens10}, thermodynamic Casimir effects in
slabs with free surfaces~\cite{Diehl12,Diehl14}, the $d = 2$,
4-state Potts model~\cite{Salas97,Shchur09,Berche13}, percolation and the 6-d Potts model \cite{PhysRevE.68.036129}, and many other systems. 
Each such system has hitherto been treated as a special case. 

Here we use the fact that the predictions of the RG can be written down as a set of differential equations in the abstract space of Hamiltonians. This allows us to apply a branch of dynamical systems theory, normal form theory~\cite{murdock2006normal, PNFT1} to provide a unified description applicable to all of these systems. We arrange these systems into {\em universality
families} of theories, each defined by its normal form. Each family has 
{\em universal terms} (linear and nonlinear), whose values 
determine a system's universality
class within the family. Finally, each family's normal form predicts the
natural {\em invariant scaling
combinations} governing universal scaling functions.


The perspective we present here is transformative: unifying, simplifying,
and systematizing a previously technical subject and promising new developments
in the field. Our best analogy is to the introduction of homotopy theory in the
1970's~\cite{toulouse1976principles,rogula1976large,Merm79,GoldbartK19} 
as a systematic method that unified the treatment of some of the many defect
structures studied in materials and field theories. Just as there have
been several previous works that correctly identified the universal effects
of nonlinear terms for phase transitions where an analytic RG 
approach is available~\cite{Wegner72,Meinke05,sonoda,magradze,pelissetto2013renormalization, barma1984corrections, barma1985two, hasenbusch2008universal, Salas97}, the 
Burger's vector, winding number, and wrapping number of dislocations in 
crystals, disclinations in liquid crystals, and Skyrmions in nuclei were 
understood individually before the mathematics of homotopy theory was seen
as the natural mathematical framework. Just as homotopy theory facilitated the
study of defects in more complex systems (metallic glasses, cosmic strings,
quasicrystals), so our normal
form methods are allowing the correct identification
and characterization of the singularity in systems in experimental and 
numerical explorations where analytic RG calculations do not yet
exist~\cite{Lorien17}. Finally, homotopy theory quickly uncovered
the fascinating entanglement and transformation properties of non-abelian
defects~\cite{Merm79}, with early speculative applications in glass
physics~\cite{Nelson83} and eventually 
inspiring the closely related nonabelian braiding being developed for 
topological quantum computing. Similarly, we demonstrate here that our methods
allow, for what appears to be the first time, the use of the correct,
remarkably rich, invariant scaling variables in the universal scaling
functions for systems where universal nonlinear RG terms are needed, and
we have discussed elsewhere~\cite{raju2018reexamining} how our methods can
be powerful tools for systematically incorporating corrections to
scaling near critical points even when universal nonlinear terms are not
needed. For example, in the future the normal-form change of variables 
we introduce here could become an inner expansion matched to 
series and virial expansions at extremes of the phase diagram; this would
allow rapid and accurate convergent characterizations of materials systems
close to and far from criticality.

Our machinery provides a
straightforward method to determine the complete form of the critical
singularity in these challenging cases. Our initial results are complex and
interesting; they pose challenges which we propose to address in future
work. The coordinate transformation to the normal form embodies analytic corrections to scaling, which allow us to address experimental
systems as they vary farther from the critical point. Finally, bifurcation theory
is designed to analyze low-dimensional dynamical systems without detailed understanding of the underlying equations;
our methods should improve scaling collapses in critical phenomena
like 2-d jamming~\cite{goodrich2014jamming}
where there is numerical evidence for logarithms but no RG framework is available.

We begin by distinguishing our work from previous literature connecting the RG to normal form theory. The previous approach~\cite{deville2008analysis, ei2000renormalization, ziane2000certain} compared the application of RG-like methods and normal form theory to solving nonlinear differential equations using perturbation theory. The connection we are making is different. We are applying normal form theory to the RG flow equations. Hence, our approach is to apply normal form theory to make predictions about the general structure of the flows given the topology (nature and number of fixed points), rather than to apply it to the model that produces these flows. 

We give an introduction to normal form theory in Section~\ref{sec:normalform}. We give a survey of the previous literature on nonlinear scaling in the RG in section~\ref{sec:earlierwork}. We show how the application of normal form theory allows us to define universality families of fixed points in Section~\ref{sec:universality}. We present several worked out examples starting with the 4-d Ising model in Section~\ref{sec:4dising} and the Random Field Ising model in Section~\ref{sec:randomfieldising}. We then work out the application of normal form theory to the Ising model in dimensions $1$, $2$ and $3$ in Sections~\ref{sec:3dising}--~\ref{sec:2dising}.

\section{Normal Form Theory}\label{sec:normalform}

Normal form theory~\cite{PNFT1} is a technique to reduce differential equations to a `normal form' by change of coordinates, often the simplest possible form. This is achieved by making near-identity coordinate transformations to get rid of as many terms as possible from the equation. It was developed initially by Poincar\'{e} to integrate nonlinear systems~\cite{poincare, chenciner2015poincare}. The physical behavior should be invariant under analytic changes of coordinates, and the length (or time) parameter should stay the same, 
which the mathematical literature addresses by perturbative polynomial changes
of coordinates (attempting removal of $n$th order nonlinearities in the flow
by using $n$th order or lower terms in the change of variables). To any 
finite order
this gives an analytic change of coordinates, but it is not in general
guaranteed to converge to an analytic transformation; we will thus
call it a polynomial change of coordinates. 

We give a brief introduction to normal form theory here for completeness. A more detailed treatment can be found in Ref.~\cite{PNFT1}. Typically one starts with a set of differential equations of the form 
\begin{equation}
 \frac{d \bm{\theta}}{d \ell} = \bm{g}(\bm{\theta}, \epsilon) ,
 \end{equation}
 where $\epsilon$ is some parameter, $\bm{\theta} = \{\theta_i\}$ is the vector of state variables and the vector field $\bm{g}$ defines the flow. In the context of statistical mechanics and renormalization group flows, $\theta_i$'s are parameters or coupling constants that enter into the free energy and $\epsilon$ is the difference in dimension from the lower or upper critical dimensions. Let us first work with the case where $\epsilon$ does not enter into the equations. The first step is to find the fixed point of the equation and use translations to set the fixed point of each $\theta_i$ is at 0. The next step is to linearize about the fixed point and reduce the linear part to the simplest possible form. In general, this is the Jordan canonical form but it is often just the eigenbasis. Then, the equation is 
 \begin{equation}
 \frac{d \bm{\theta}}{d \ell} = J \bm{\theta} + \bm{f}(\bm{\theta}) ,
\end{equation}
where $J$ is the linearized matrix of the flow and the remaining terms are in the vector field $\bm{f}(\bm{\theta}) \sim \mathcal{O}(\theta^2)$. Terms of order $k$ are defined to be made up of homogeneous polynomials of order $k$. So for $\bm{\theta} = (\theta_1, \theta_2, \theta_3)$, $\theta_1^2 \theta_2 \theta_3 \sim \mathcal{O}(\theta^4)$. We will denote terms of order $k$ by a lower index. So 
\begin{equation}
 \bm{f}(\bm{\theta}) = \sum_{k \geq 2} \bm{f}_k (\bm{\theta}) .
\end{equation}
Note the index is giving the order of the polynomial and not enumerating the components of the vector field. Let the lowest non-zero term be at some order $k \geq 2$ (usually 2). Then we can write
\begin{equation}
 \frac{d \bm{\theta}}{d \ell} = J \bm{\theta} + \bm{f}_k (\bm{\theta}) + \mathcal{O}(\theta^{k+1}) .
 \end{equation}
The idea is to try and remove higher order terms by making coordinate changes. To remove the term $\bm{f}_k$, we try to do a coordinate change of order $k$,  
\begin{equation}
\label{changecoord}
 \bm{\theta} = \bm{\tilde \theta} + \bm{h}_k (\bm{\tilde \theta}) ,
\end{equation}
where $\bm{h}_k (\bm(\tilde \theta))$ is a polynomial in $\bm{\tilde \theta}$. This construction is similar  to nonlinear scaling fields~\cite{cardy1996scaling, Wegner72} which try to linearize the RG flow equations with a subtle difference that we will remark on later. The higher order terms which we can remove by coordinate changes correspond to analytic corrections to scaling. Then, to find the equations in the new variables.
\begin{equation}
 \frac{d \bm{\theta}}{d \ell} = \frac{d \bm{\tilde \theta}}{d \ell} + (\mathcal{D} \bm{h}_k) \frac{d \bm{\tilde \theta}}{d\ell} .
\end{equation}
$\mathcal{D} \bm{h}_k$ is the matrix of partial derivatives of the vector field $\bm{h_k}$ with respect to the parameters $\bm{\theta}$. Now, substituting this into the equation
\begin{equation}
 (1 + \mathcal{D} \bm{h}_k) \frac{d \bm{\tilde \theta}}{d \ell} = J (\bm{\tilde
    \theta} + \bm{h}_k) + \bm{f}(\bm{\tilde \theta} + \bm{h}(\bm{\tilde
    \theta})) + \mathcal{O}({\tilde \theta}^{k+1}), 
 \end{equation}
which upon simplification gives
 \begin{equation}
 \frac{d \bm{\tilde \theta}}{d \ell} = J \bm{\tilde \theta} - (\mathcal{D} \bm{h}_k) J \bm{\tilde \theta} + (\mathcal{D} J \bm{\tilde \theta}) \bm{h}_k + \bm{f}_k (\bm{\tilde \theta})+ \mathcal{O}({\tilde \theta}^{k+1}) .
\end{equation}
For the last line, notice that the matrix $J$ is the same as $\mathcal{D} J \bm{\tilde \theta}$ (i.e. the same as the matrix of partial derivatives with respect to parameters $\bm{\tilde \theta}$ of the vector $J \bm{\tilde \theta}$). 
Two of the terms can be written as the Lie bracket (a commutator for vector fields) defined as $[\bm{h}_k, J \bm{\tilde \theta}] = - ( (\mathcal{D} \bm{h}_k) J \bm{\tilde \theta} - (\mathcal{D} J \bm{\tilde \theta}) \bm{h}_k)$  to give the final equation 
\begin{equation}
 \frac{d \bm{\tilde \theta}}{d \ell} = J \bm{\tilde \theta} + [\bm{h}_k, J \bm{\tilde \theta}] + \bm{f}_k + \mathcal{O}({\tilde \theta}^{k+1}) .
\end{equation}
So, if we want to remove the term $\bm{f}_k$, we need to solve the equation  $[\bm{h}_k, J \bm{\tilde \theta}] =  -\bm{f}_k$ for $\bm{h}_k$. It's important to note that whether this equation can be solved or not depends only on the linear part of the equation given by the matrix J. That is, within the space of transformations that we are considering, the linear part of the equation completely determines how much the equation can be simplified and how many terms can be removed. This is not true if there are zero eigenvalues and one then has to consider a broader space of transformations which we will consider later.

To see when the equation can be solved, we first note that the space of homogeneous polynomials is a vector space with a basis constructed in the obvious way $\theta_1^{\alpha_1}...\theta_n^{\alpha_n}$. Any term at order $k$ can be written as a sum of such terms for which $\sum_i \alpha_i = k$. The Lie bracket can be thought of as a linear operator on this space. To find the set of possible solutions is to find the range of this linear operator. Let us take the case where the linear part is diagonalizable and so just consists of the eigenvalues $\lambda_i$. Let us say for simplicity that the $j$ component of the vector $\bm{f_k}$ $(f_k)^j = c {\tilde \theta}_1^{\alpha_1}...{\tilde \theta}_n^{\alpha_n}$ for some set of \{$\alpha_i$\}. Then, the $j$th component of the matrix equation reduces to


\begin{equation}
 \lambda_j (h_k)^j - \left(\sum_i \lambda_i \alpha_i\right) (h_k)^j = c \theta_1^{\alpha_1}...\theta_n^{\alpha_n} .
\end{equation}
This can be solved easily by choosing $(h_k)^j = a {\tilde \theta}_1^{\alpha_1}...{\tilde \theta}_n^{\alpha_n}$ and
\begin{equation}
\label{normalformequation}
 a = \frac{c}{\lambda_j - \sum_i \lambda_i \alpha_i} .
\end{equation}
When all nonlinear terms can be removed by such a coordinate transformation, then the usual case of power law scaling is obtained. The fixed point, in this case, is called hyperbolic. Alternatively, if we have a term with  $\lambda_j = \sum_i \lambda_i \alpha_i$ (a linear combination of other eigenvalues $\lambda_i$ with positive integer coefficients $\alpha_i$), this term is called a resonance and cannot be removed from the equation for $d \theta_j/d \ell$. This contributes to the singularity at the fixed point which is no longer given by power law combinations. 

\subsection{Bifurcations}

Notice a special case of these equations when for some $k$, a particular $\lambda_k = 0$. In this case, it is possible to get an infinite number of resonances because the equation $\lambda_i = \lambda_i + \alpha_k \lambda_k$ is also true for all $\alpha_k$ and $\lambda_i$. This case, when one of the eigenvalues goes to 0 depending on some parameter $\epsilon$ is called a \textit{bifurcation}. If all linear eigenvalues $\lambda_i$ of the flows are distinct and
non-zero, which terms can be removed using polynomial coordinate changes
depends only on these $\lambda_i$. As we saw, this approach can be formulated
elegantly as a linear algebra problem of the Lie bracket on the space of
homogeneous polynomials.  For more
general cases---including bifurcations---`hypernormal form'~\cite{murdock2004hypernormal, yu2007simplest,
yu2002computation} theory develops a systematic but somewhat more
brute-force machinery to identify which terms can and cannot be removed
perturbatively by polynomial changes of coordinates. Classic bifurcations include the pitchfork bifurcation, the transcritical bifurcation, the saddle node and the Hopf bifurcation~\cite{GuckenheimerH13}. 

Confusingly, bifurcation theory separately has its own `normal form' of bifurcations. These normal forms are derived in a very different way using the implicit function theorem. The basic idea is to ask for the smallest number of terms in the equation which will preserve the qualitative behavior of the fixed points (e.g. exchange of stability of fixed points), and then map any other equation on to this simple equation using the implicit function theorem. This mapping allows for too broad a class of transformations to be useful for our purposes. An important feature of the flows that we want to preserve is their \textit{analyticity}, we therefore only consider polynomial changes of coordinates.

An explicit example is given by the 4-d Ising model. It is known that the magnetization $M \sim t^{1/2} (\log t)^{1/3}$ with $\log \log$ corrections. The quartic coupling $u$ and the temperature $t$ have flow equations which traditional bifurcation theory would simplify to
\begin{align}
 \frac{d u}{d \ell} &= - \bar{B} u^2 , \\
 \frac{d t}{d \ell} &= 2 t .
\end{align}
Calculating the magnetization with this set of flow equations leads to the wrong power of logarithmic corrections. By allowing too broad a class of coordinate transformations, bifurcation theory hides the true singularity in the non-analytic coordinate change. We will show that normal form theory instead predicts  
\begin{align}
 \frac{d u}{d \ell} &= -\bar{B} u^2 + \bar{D} u^3 , \\
 \frac{d t}{d \ell} &= 2 t - \bar{A} t u ,
\end{align}
which does predict the correct behavior. We will present the explicit solution of this equation in Section~\ref{sec:4dising}. Here, we just note that the traditional $\log$ and $\log \log$ terms follow from the solution's  asymptotic behavior.  To get these equations, we will remove higher order terms in $u$ by using a coordinate change that is lower in order (broadening the formalism we considered in Section~\ref{sec:normalform}). Using lower order terms to remove higher order terms is part of hypernormal form theory. For our purposes, the distinction is somewhat artificial and  here we simply use normal form theory to denote any procedure that uses only polynomial changes of coordinates to change terms in flow equations. 


In Sections~\ref{sec:4dising} and~\ref{sec:randomfieldising}, we will explicitly work out the case of a single variable undergoing a bifurcation for the 4d Ising model and the 2d Random Field Ising model and show how there are only a finite number of terms which cannot be changed or removed. It is worth mentioning here that there can be cases in which two variables simultaneously have 0 eigenvalues. The XY model~\cite{kosterlitz1974critical} offers an example where this happens. The dATG transition in 6 dimensions has two variables that simultaneously go through a transcritical bifurcation~\cite{charbonneau2017nontrivial, Yaida18}. Polynomial changes of coordinates in both variables can be used here too, but because there are generically more terms at higher order than at lower order (there are many more ways to combine two variables into a sixth order polynomial than there are to combine them into a third order polynomial), we usually do not have enough freedom to remove all terms. Therefore, simultaneous bifurcations in more than one variable often have an infinite number of terms in their flow equations that cannot be removed.

\section{Earlier work} \label{sec:earlierwork}

The approach we take is inspired by Wegner's early work~\cite{Wegner72,
Wegner73}, subsequent developments by Aharony and
Fisher~\cite{aharony1980university, aharony1983nonlinear}, and by studies of Barma and Fisher on
logarithmic corrections to scaling~\cite{barma1984corrections,
barma1985two}. The approach of Salas and Sokal on the 2-d Potts model~\cite{Salas97}, and  of Hasenbusch~\cite{hasenbusch2008universal} et al. on the 2d Edwards-Anderson model 
is similar in spirit to ours.

Wegner~\cite{Wegner72} first constructed nonlinear scaling fields which transform linearly under an arbitrary renormalization group. His construction is very similar to the coordinate changes we considered above for normal form theory. The one difference is that Wegner explicitly allows the new coordinates to depend on the coarse graining length $\ell$. We will not allow this explicit dependence on $\ell$ in our change of coordinates, as it doesn't seem to offer any advantage over regular normal form theory.




Eventually, the goal of using normal form theory to understand the differential equations that describe RG flow is to simplify and systematize scaling collapses. This requires a systematic way of dealing with corrections to scaling beyond the usual power laws. There are three different types of corrections to scaling that have appeared in the literature. These include logarithmic, singular and analytic corrections to scaling. Logarithms in the scaling behavior typically occur at an upper critical dimension or in the presence of a resonance. Wegner and Riedel~\cite{Wegner73} considered the case of a zero eigenvalues which occurs at the upper critical dimension of Ising and tri-critical Ising models. They derived the form of the scaling in terms of logarithmic corrections to scaling. However, they used perturbation theory to ignore higher order terms in the flow equations rather than only keeping those terms which cannot be removed by an application of normal form theory. Here, we will solve the full flow equations and see that the logarithmic corrections to scaling are better incorporated as part of the true singularity using normal form theory.

Analytic corrections to scaling were explored by Aharony and Fisher~\cite{aharony1983nonlinear} who gave a physical interpretation of the nonlinear scaling fields (see below Eq.(~\ref{changecoord})) in terms of analytic corrections to scaling in the Ising model. Analytic corrections to scaling capture the difference between the physical variable $T$ and $H$ (that your thermometer or gaussmeter measures) and the symbols $\tilde{t}$ and $\tilde{h}$ in the theory of the magnet. The liquid gas transition is in the Ising universality class but a theory of the liquid gas transition has to include analytic corrections to scaling to match with the universal predictions of the Ising model. Moreover, such corrections are also needed to explain the non-universal behavior away from the fixed point. Analytic corrections to scaling will correspond to terms in the differential equations that can be removed by coordinate changes.

The singular corrections to scaling are also incorporated as part of the true singularity with the addition of irrelevant variables.  Finally, the ability to change the renormalization scheme leads to what are called redundant variables. In related work~\cite{raju2018reexamining}, we argue that these variables can be seen as a gauge choice which contributes to the corrections of scaling. In forthcoming work~\cite{Clement18}, we will explore the consequence of this distinction between gauge corrections and genuine singular corrections to scaling further.

Finally Salas and Sokal, in the context of the 2-d Potts model, derive the normal form of the flow equations for a transcritical bifurcation. Similarly, Hasenbusch et al. derive the normal form for the 2-d Edwards-Anderson model which is also a transcritical bifurcation. Both of these do not solve the full flow equations but end up approximating the solution by logarithms. In the context of QCD, Sonoda derived the solution for the flow of a coupling which undergoes a transcritical bifurcation. 

Despite similar inclinations, none of these works make the complete
connection to normal form theory. One advantage of our approach is precisely that it brings together this disparate literature into a unified framework.  The
analysis presented here is general and applicable to all kinds of
situations, ranging from old problems like the nonequilibrium random field Ising
model (NERFIM)~\cite{PerkovicDS95}, to newer research problems like jamming~\cite{goodrich2014jamming}.

\section{Universality Families} \label{sec:universality}

\begin{table*}
\begin{center}
\resizebox{\linewidth}{!}{
  \begin{tabular}{ | l | c | c| c| }
    \hline
    Universality family & Systems & Normal form & Invariant scaling combinations \\ \hline
     \makecell{\raisebox{-0.4\height}{\includegraphics[width=.06\textwidth]{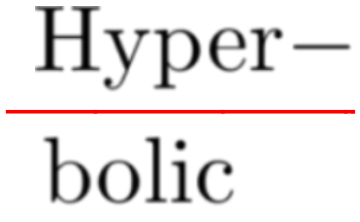}}}  
     
    & \makecell{\textbf{3-d Ising Model $(t)$} \\ 3-d RFIM $(w)$} & $dt/dl = (1/\nu) t$ &

		 $L t^{\nu}$ \\ \hline
    \makecell{\raisebox{-0.4\height}{\includegraphics[width=.06\textwidth]{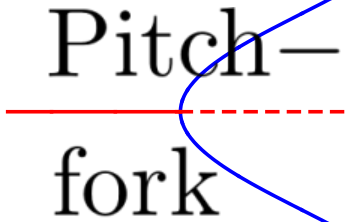}}}
    & \makecell{\textbf{2-d RFIM $(w)$} \\ 6-d Potts model $(q)$} & $dw/dl =  w^3 + \textcolor{blue}{B w^5}$ & 

		$L e^{1/(2 w^2)} 
		(\textcolor{darkGreen}{1/w^2} \textcolor{blue}{+ B})^{\textcolor{blue}{-B/2}}$\\ \hline
     \makecell{\raisebox{-0.4\height}{\includegraphics[width=.09\textwidth]{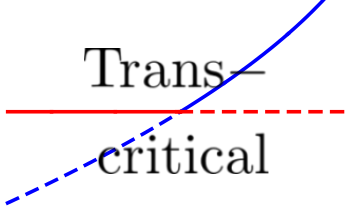}}}
    & \makecell{ \textbf{4-d Ising model $(u,t)$} \\ 2-d NERFIM $(-w,S)$ \\ 1-d Ising model $(-t,h)$ }  & \makecell{$du/dl = -u^2 + \textcolor{darkGreen}{D u^3}$ \\ $dt/dl = 2 t \textcolor{darkGreen}{- A t u}$}  & 
	\makecell{$L e^{1/u - D} \textcolor{blue}{(1/(D u) - 1)^{D}} = L y^{D}$  \\
	 $t L^2\textcolor{blue}{(W(y L^{1/D})/(1/(D u) - 1))^{-A} }$}
       \\ \hline
      \makecell{Resonance} & \textbf{2-d Ising model} 
    & \makecell{$df/dl = 2 f \textcolor{darkGreen}{- t^2} \textcolor{darkGreen}{- L^{-2}}$ \\ $dt/dl = t\textcolor{darkGreen}{+AL^{-1}}$} & 
	$tL \textcolor{darkGreen}{+ A\log L}$ \\ \hline
    \makecell{\raisebox{-0.4\height}{\includegraphics[width=.1\textwidth]{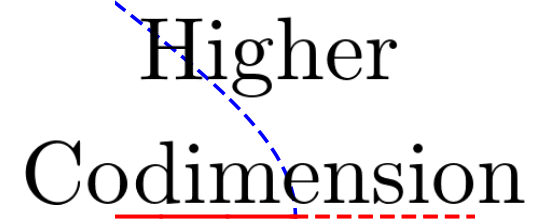}}}
    & \textbf{2-d XY model} &
    \makecell{$dx/dl=-y^2(1\textcolor{darkGreen}{+xf(x^2)})$ \\
  $dy/dl = -xy$} &
   \makecell{$y^2 -2\int_0^x s/(1\textcolor{darkGreen}{+sf(s^2)})\,ds$\\ 
	$=y^2-x^2\textcolor{darkGreen}{-(2f(0)/3)x^3+(f(0)^2/2)x^4+\mathcal
  O(x^5)} $}\\ \hline
 
  \end{tabular}}
  \end{center}
  \caption{Normal forms and universal invariant scaling combinations for 
traditional and intrinsically nonlinear renormalization-group critical points.
The universal scaling of most critical points are power-law combinations
of the control variables, derived from the linearized normal-form equations
of hyperbolic RG fixed points. Many systems have well-studied
logarithmic corrections, exponentially diverging correlations, or other
singularities that we attribute to intrinsic nonlinearities in the 
RG flow equations. 
In blue are new universal terms predicted by our analysis of the
corresponding dynamical system normal forms, which appear not to have
been hitherto discussed in the literature.
In green are terms we explain which have been previously observed using other
methods~\cite{Wegner72,Meinke05,sonoda,magradze,pelissetto2013renormalization}. The normal form equations are shown for the system in bold. Other systems in the same universality family have the same equations associated with different variables (shown in parenthesis). The invariant scaling combination for the transcritical family requires the Lambert $W$ function defined by the equation $W(x) \exp(W(x)) = x$. Many of the results quoted in the table were obtained in disparate literatures (QCD, glasses, critical phenomena etc.) but are united in this common framework. Other families are possible, the flow equations for the replica symmetry breaking transition in disordered media have a simultaneous transcritical bifurcation and possibly also have a Hopf bifurcation~\cite{Yaida18}} .
\label{systemstable}
\end{table*}

Traditionally, the RG contains the concept of a universality class. The universality class is essentially determined by the critical exponents which explain the scaling behavior of a model, i.e. by linearized RG eigenvalues. Normal form theory suggests another possible classification. Each fixed point can be classified by the bifurcation or resonance that it is at. The simplest case, which is also the traditional one, is the hyperbolic universality family. In the hyperbolic case, it is possible to remove all nonlinear terms in the flow equations by changes of coordinates. Hence, the RG can be written as a linear flow to all orders in perturbation theory. Different values for the linear eigenvalues correspond to different universality classes. While traditionally this is a statement about the linearization of the RG, here it is a statement about the only terms in the flow equations that are \textit{universal} in the sense that they can not be removed by a coordinate change. 

The need for this generalization becomes clear when we examine cases which are not traditional. In  Table~\ref{systemstable} we present common universality families 
and well-studied statistical mechanics systems governed by each.
The pitchfork bifurcation shows up
in the 2-d Random Field Ising model; it has a cubic term in the equations
for $w$, the ratio of the disorder to the coupling~\cite{Bray85}. We
have derived that the correct equations require an additional $w^5$
term~\cite{Lorien17}, which was not included in previous work. The 2-d
Ising model has a well known logarithmic correction to the specific
heat, which Wegner associated with a $t^2$ resonance term in the flow
equation~\cite{Wegner72}. The 1-d and 4-d Ising models have transcritical
bifurcations. The 1-d Ising case is somewhat special and we will cover it later in Section~\ref{sec:1dising}. These cover all the important bifurcations with one variable~\footnote{We have not studied any example of a saddle node bifurcation which would require a transition from a critical point to no critical point.}.

When more than one variable is undergoing a bifurcation, or if more than one variable has an inherently nonlinear flow, the analysis becomes considerably more complicated. This is evidenced in the the 2-d XY model at the Kosterlitz--Thouless (KT) transition~\cite{kosterlitz1977d}.  It has been shown that the simplest normal form
of its flow equations (in the inverse-temperature-like variable
$x\sim1/T-1/T_c$ and the fugacity $y$) has an infinite number of
universal terms, which can be rearranged into an analytic
function $f$~\cite{pelissetto2013renormalization} (Table~\ref{systemstable}).
We conjecture that the very similar transition observed in randomly grown
networks~\cite{callaway2001randomly,dorogovtsev2001anomalous} is not
in the KT-universality class, but rather is in the same universality
family. It is not to be expected that a percolation transition for
infinite-dimensional networks should flow to the same fixed point
as a 2-d magnetic system, but it is entirely plausible that they share
the same normal form with a different universal function $f$.

Different universality classes within the same universality family, such as those of the 4-d Ising model and the 2-d NERFIM have different power laws and scaling functions. However, as shown in Table~\ref{systemstable}, because they both have a transcritical bifurcation the two classes have the same complicated invariant scaling combinations~\footnote{A correlation length $y^{-D}$ from Table~\ref{systemstable} defined in terms of the marginal variable in both cases diverges exponentially; in terms of the temperature the correlation length is a power law}. This hidden connection is made apparent in the shared normal form, where the quartic coupling and temperature ($u,T$) in the first class are associated with the (negative of) disorder strength and avalanche size ($-w,S$) in the second~\footnote{The minus sign on $w$ and $t$ for the 1-d Ising
and the NERFIM is because $w$ and $t$ are 
marginally relevant whereas $u$ is marginally irrelevant for 4-d Ising.}.


Indeed, the normal form not only unites these universality classes, but allows a more precise handling of their singularity. It is usually stated that the magnetization $M \sim t^{1/2} (\log
t)^{1/3}$, the specific heat $C \sim (\log t)^{1/3}$ and the
susceptibility $\chi \sim (\log t)^{1/3}/t$ with $\log \log$ corrections~\cite{Wegner73}. We show in the supplementary material that
the true singularity of the magnetization
at the critical point is $M \sim t^{1/2} W(x t^{-27/25})^{1/3}$,
where $W$ is the Lambert-W function defined by $W(z) e^{W(z)} = z$, and
$x(u)$ is a complicated but explicit function of the irrelevant variable $u$.
(The traditional log
and log-log terms follow from the asymptotic behaviors of $W(x)$ at large
and small $x$. The universal power $27/25$ becomes manifest in the 
complete singularity, but is disguised into a constant factor up to
leading logs.) We now show how to apply normal form theory to specific examples.

\section{Application to specific systems}
In the sections below, we derive in detail the scaling form for the entries shown in Table~\ref{systemstable}. Our archetypal example is the 4-d Ising model. For this, we derive the scaling forms and use them to perform scaling collapses of numerical simulations. We then discuss the scaling of the Random Field Ising model, the XY model and the Ising model in dimensions $1$, $3$ and $2$. 

\subsection{4-d Ising} \label{sec:4dising}

The study of critical points using the renormalization group was turned into a dynamical system problem by Wilson~\cite{wilson1974renormalization}. These RG calculations are done by first expressing the Ising model as a field theory with a quartic potential $u \phi^4$. Then by coarse-graining in momentum space and rescaling, one obtains the flow equations
\begin{align}
 {d t}/{d \ell} =& 2 t -  \bar{A} t u + \bar{C} t u^2 + \bar{E} t u^3 \nonumber \\ &+ \bar{G} t u^4 + \bar{I} t u^5 + \bar{K} t u^6 ... , \\
 {d u}/{d \ell} =& \epsilon u - \bar{B} u^2 + \bar{D} u^3 + \bar{F} u^4 \nonumber \\ &+ \bar{H} u^5 + \bar{J} u^6 + \bar{L} u^6  ... , \\
 {d f}/{d \ell} =& (4 - \epsilon) f + ...,
 \label{wilsoneqs}
\end{align}
where $t$ is the temperature, $f$ is the free energy and $u$ is the leading irrelevant variable. This is the highest order to which the flow equations are known as of now. The coefficients take the values, $\bar{A} = 1$, $\bar{B} = 3$, $\bar{C} = 5/6$, $\bar{D} = 17/3$, $\bar{E} = -7/2$, $\bar{F} \approx 32.54$, $\bar{G} \approx 19.96$, $\bar{H} \approx -271.6$, $\bar{I} \approx -150.8$, $\bar{J} \approx 2849$, $\bar{K} \approx 1355$, $\bar{L} \approx -34776$~\cite{kompaniets2016renormalization, chetyrkin1983five}. The flow equation for $u$ in this case takes the form of a transcritical bifurcation with parameter $\epsilon = 4 - d$ tuning the exchange of stability between the Gaussian ($u = 0$) and Wilson-Fisher fixed point ($u \neq 0$). 

Consider these equations for $\epsilon = 0$ which is the point at which it undergoes a transcritical bifurcation. To derive the normal form, one considers a change of variables of the form 
\begin{align}
 t &= \tilde t + a_1 \tilde t \tilde u + a_2 \tilde t \tilde u^2 + ... , \\
 u &= \tilde u + b_1 \tilde u^2 + b_2 \tilde u^3 + b_3 u^4 + ...
\end{align}
This gives the equations up to order $u^4$, 
\begin{align}
 {d \tilde t}/{d \ell} =& 2 \tilde t -  \bar{A} \tilde t \tilde u + (-\bar{A} b_1 + a_1 \bar{B} + \bar{C}) t u^2 + ... , \\
 {d \tilde u}/{d \ell} =& - \bar{B} \tilde{u}^2 + \bar{D} \tilde{u}^3 \nonumber \\ &+ (-b_1^2 \bar{B} + b_2 \bar{B} + b_1 \bar{D} + \bar{E}) \tilde{u}^4 + ...
\end{align}
Note that any term of the form $u^m t$ in the equations for $d t/d \ell$ and any term of the form $u^m$ in the equations for $d u/d \ell$ is a resonance. Hence, the coefficients $\bar{A}$, $\bar{B}$ and $\bar{D}$ remain unchanged with this change of variables. However, the coefficients $\bar{C}$ and $\bar{E}$ are changed (though the change is independent of $a_2$ and $b_3$ because they are resonances) and in particular, can be set to 0 by an appropriate choice of coefficients. 

This creates a general procedure for reducing this flow to its simplest possible form. First, all terms that are not resonances are removed in the usual way by solving Eq.(~\ref{normalformequation}). Then, we perturbatively remove most of the resonances using the following procedure. First consider the $u$ flow. Suppose the lowest order term in the flow after the $u^3$ term is $u^n$, i.e.

\begin{equation}
 \frac{d u}{d \ell} = - \bar{B} u^2 + \bar{D} u^3 + \bar{N_n} u^n + \mathcal{O}(u^{n+1})
\end{equation}
with $n > 3$. Consider a change of variables of the form $u = \tilde{u} + b_{n-2} \tilde{u}^{n-1}$. Then
\begin{widetext}
\begin{align}
 (1 + (n-1) b_{n-2} \tilde{u}^{n-2}) \frac{d \tilde u}{d \ell} &= - \bar{B} (\tilde{u} + b_{n-2} \tilde{u}^{n-1})^2 + \bar{D} ((\tilde{u} + b_{n-2} \tilde{u}^{n-1})^3 \nonumber \\ &+ \bar{N_n} (\tilde{u} + b_{n-2} \tilde{u}^{n-1})^n + \mathcal{O}(\tilde{u}^{n+1}) , \\
 \frac{d \tilde u}{d \ell} &= \frac{- \bar{B} \tilde{u}^2 + \bar{D} \tilde{u}^3 + \bar{N_n} \tilde{u}^n - 2 \bar{B} b_{n-2} \tilde{u}^n}{(1 + (n-1) b_{n-2} \tilde{u}^{n-2}) } + \mathcal{O}(\tilde{u}^{n+1}) , \\
 &= - \bar{B} \tilde{u}^2 + \bar{D} \tilde{u}^3 + (\bar{N_n} - 2 \bar{B} b_{n-2} + (n-1) b_{n-2} \bar{B}) \tilde{u}^n + \mathcal{O}(\tilde{u}^{n+1}) .
\end{align}
\end{widetext}
Evidently, the coefficient of the $\tilde{u}^n$ term can be set to 0 with an appropriate choice  $b_{n-2} = N_n/(\bar{B} (3 - n))$.

So all terms of the form $u^n$ with $n > 3$ can be removed by a change of coordinates. Incidentally, this derivation also shows why it is not possible to remove the $u^3$ term. Now consider the $t$ equation with
\begin{equation}
 \frac{dt}{d \ell} = 2 t - \bar{A} t \tilde u + M_n t \tilde u^{n-1} + \mathcal{O} (t \tilde u^{n}) .
\end{equation}
We consider a change of coordinates
\begin{equation}
 t = \tilde t + a_{n-2} \tilde t \tilde{u}^{n-2} .
\end{equation}
It is then straightforward to show
\begin{equation}
 \frac{d \tilde t}{d \ell} = 2 \tilde t - \bar{A} \tilde t \tilde u + (M_n + \bar{B} (n-2) a_{n-2} + a_{n-2} \bar{A}) \tilde t \tilde u^{n-1} + \mathcal{O} (t \tilde u^{n}) .
\end{equation}
So setting $a_{n-2} = - M_n/(\bar{B}(n-2) + A)$ sets the coefficient of the $t u^{n-1}$ term with $n > 2$ to 0. 

Any term which is not of this form can be removed in the usual way by solving Eq.~\ref{normalformequation}. Finally, we have another degree of freedom that we have used. We can rescale $u$ and $t$ to set some of the nonlinear coefficients to 1. This reflects the fact that the original coefficients $\bar{A}$, $\bar{D}$ depend on an arbitrary scale of $u$ and $t$ that we have chosen. By choosing the scale so $\bar{B} = 1$ and the coefficient of the resonance is -1, defines $D = \bar{D}/\bar{B}^2$ and $A = \bar{A}/\bar{B}$. Hence, by considering all such polynomial change of coordinates, we can reduce this set of equations to their normal form



\begin{align}
 {d \tilde t}/{d \ell} &= 2 \tilde{t} - A \tilde{u} \tilde{t} , \label{isinghnf} \\ 
 {d \tilde u}/{d \ell} &= - \tilde{u}^2 + D \tilde{u}^3, \label{isinghnf2} \\
 {d \tilde f}/{d \ell} &= 4 \tilde f - {\tilde t}^2 \label{isinghnf3} .
\end{align}

The resultant equations then have 2 parameters $A$ and $D$ which
are \textit{universal}, in a way that is similar to the eigenvalues of the RG flows as in Table~\ref{systemstable}. The normal form variables $\tilde t$, $\tilde u$, $\tilde f$ are equal to the physical variables $t$, $u$ and $f$ to linear order (up to a rescaling). Corrections to these are analytic corrections to scaling. Hence, we will henceforth simply refer to the normal form variables as $t$, $u$ and $f$. It is important to note that we are making a particular choice for the analytic corrections to scaling by setting them equal to zero. It is possible to make a different choice for the higher order coefficients. In particular, the equation for $d u/d \ell$ goes to $\infty$ at finite $\ell$ if $u$ starts at a large enough value. Hence, it may be more useful to make a different choice for the higher order coefficients. All of these choices will agree close to the critical point but will have different behavior away form the critical point. Later, we will consider a different choice for the higher order terms.

The 4-d Ising model has both a bifurcation and a resonance. The $u^2$, $u^3$ and $A u t$ terms come from the bifurcation and cannot be removed
by an analytic change of coordinates. The $t^2$ term is a consequence of
an integer resonance between the temperature and free energy eigenvalue,
$\lambda_t = 1/\nu = 2, \lambda_f = d = 4$.

\begin{figure}
 \includegraphics[width=.85\linewidth]{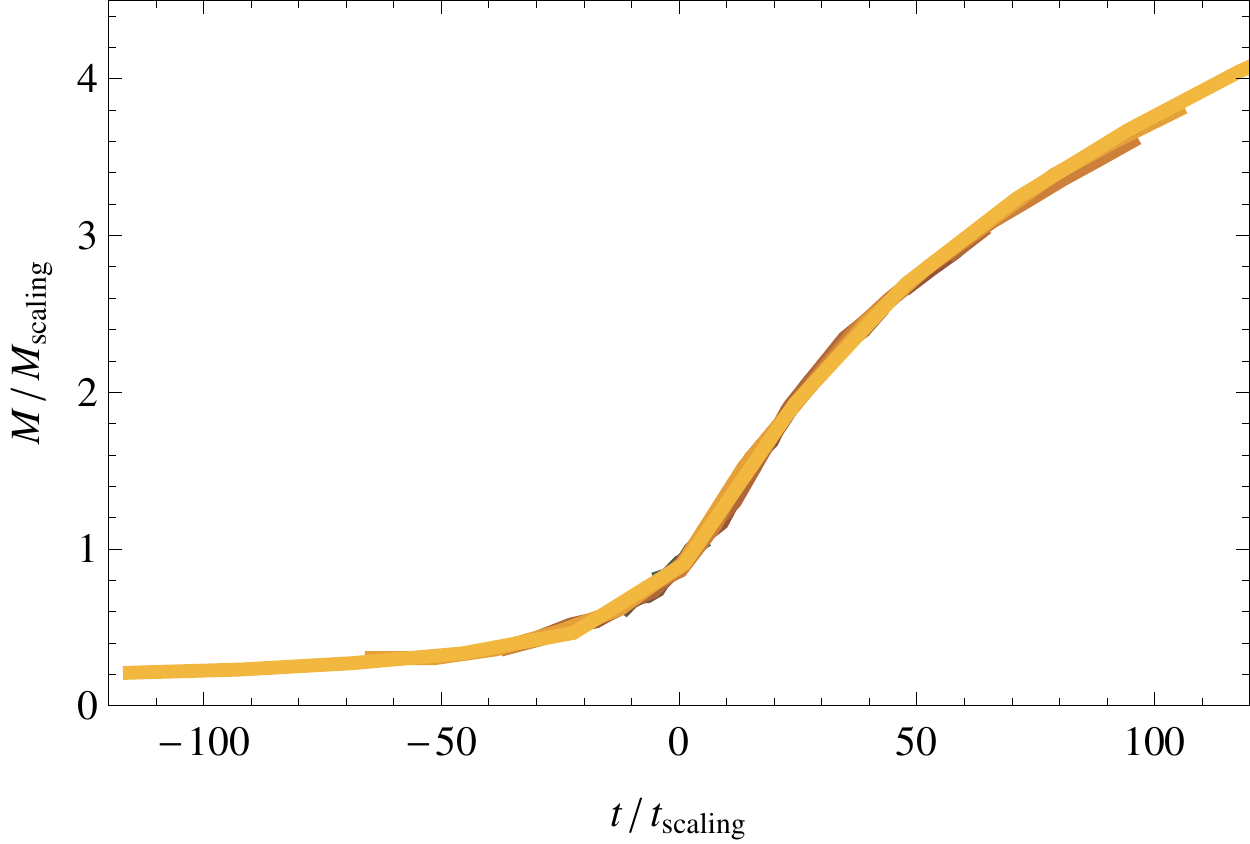}
\caption{Scaling collapses for the magnetization and susceptibility using the scaling form given by the normal form Eqs.(~\ref{isinghnf}~--~\ref{isinghnf3}). Simulations are done on a 4-d lattice using a Wolff algorithm for lattice sizes ranging from $L = 4$ to $L = 32$. Here $M_\mathrm{scaling}$ $=$ $L ((W(y L^{1/D})+1))^{1/4}$ and $t_{\mathrm{scaling}}$ $=$ $L^{-2} (W(y L^{1/D})/(1/D u_0 - 1))^{1/3} ((W(y L^{1/D})+1))^{-1/2}$. We find $u_0 = 0.4\pm0.1$ for the 4-d nearest-neighbor hypercubic-lattice. An estimate of the error is given by estimating $u_0$ with a different choice of normal form which gives $u_0 = 0.5$.  }
\label{collapses1}
 \end{figure}

\begin{figure}
 \includegraphics[width=.85\linewidth]{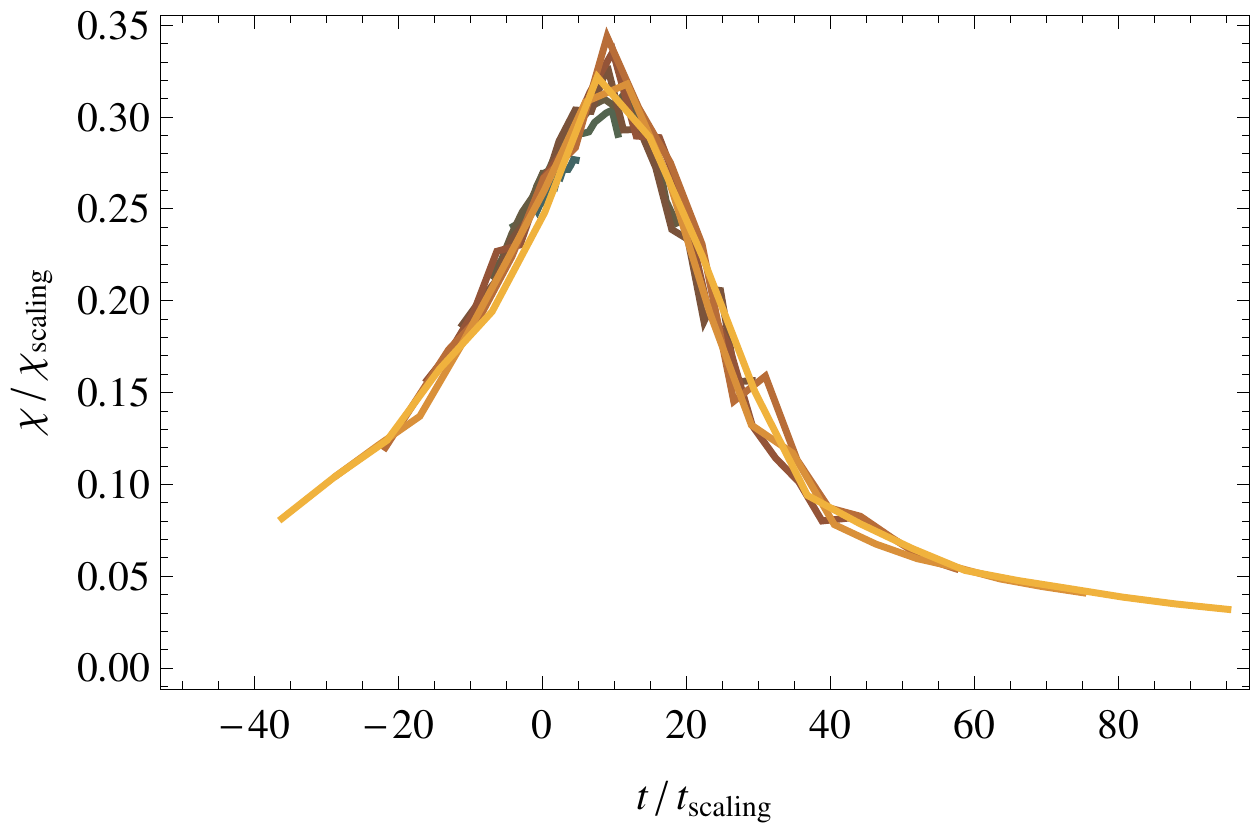}
 \caption{Scaling collapse for the susceptibility using the scaling form given by the normal form Eqs.(~\ref{isinghnf}~--~\ref{isinghnf3}). Simulations are done on a 4-d lattice using a Wolff algorithm for lattice sizes ranging from $L = 4$ to $L = 32$. Here $\chi_\mathrm{scaling}$ $=$ $L^2 ((W(y L^{1/D}) + 1))^{1/2}$ and $t_{\mathrm{scaling}}$ $=$ $L^{-2} (W(y L^{1/D})/(1/D u_0 - 1))^{1/3} ((W(y L^{1/D})+1))^{-1/2}$. We find $u_0 = 0.4$ for the 4-d nearest-neighbor hypercubic-lattice.}
\label{collapses2}
 \end{figure}



Before examining the full solution
of Eqs.~(\ref{isinghnf}~--~\ref{isinghnf3}), we will first study the effect
of each part of the RG flows. First, considering only the linear terms and
coarse-graining until $t(\ell^*) = 1$, the free energy is given by 
$f \sim t^2$. This is the mean-field result and also the traditional scaling
form that RG results take in the absence of nonlinear terms in the flow
equations. Second, we include the resonance between the temperature and
free energy eigenvalue, which leads to an irremovable $t^2$ term in
the flow equation for the free energy. This term cannot be removed by analytic
coordinate changes, and yields a $\log$ correction to the specific heat.
Third, the irrelevant variable $u$ undergoes a transcritical
bifurcation. Results in the hyper-normal form theory literature, as well
as some articles in the high-energy theory
literature~\cite{sonoda,magradze} recognize that the simplest form that 
the equation can be brought into is Eq.~\ref{isinghnf2}.  The solutions
of Eqs.(~\ref{isinghnf}~--~\ref{isinghnf2}) are $u(\ell) = 1 /(D (1 + W(y e^{\ell/D})))$ and
$t(\ell) = t_0 e^{2 \ell} (W(y e^{ \ell/D})/(1/(D u_0) - 1))^{-A}$  where $y[u_0]$ is again
a messy but explicit function: $y = (1/(D u_0)-1) \exp(1/(D u_0)-1)$. We show how to derive this in the supplementary material. 
The traditional log and log-log corrections are
derived by expanding the $W$ function for large $\ell$.

Let us use this to derive the finite-size scaling form of the free energy. Early finite-size scaling work~\cite{aktekin2001finite, lai1990finite, montvay1987numerical} attempted scaling collapses with logs; recent work does not attempt collapses at all~\cite{lundow2009critical}. Finite-size scaling requires an equation for the magnetic field, $h$, given by $dh/d\ell = 3 h$. Explicit calculations show that the coefficient of the $h u$ term is zero (see below). The free energy is then a function of three scaling variables, $u(\ell)$, $t(\ell)$ and $h(\ell)$. It is given by 
\begin{align}
    f(t_0, u_0) &= e^{-4 \ell} f(t(\ell), u(\ell), h(\ell))\nonumber\\
    &-  W(y e^{\ell/D})^{-A} \left( \frac{W(y e^{\ell/D})^{-A}}{1 - A}-\frac{1}{A}\right).
\end{align}

 To get a finite size scaling form, we coarse grain until $\ell = \log L$, the system size. Note that $u(L)$ cannot just be ignored because it is a dangerous irrelevant variable. However we can account for it by taking the combination $t(L)/(u(L))^{1/2}$ and $h(L)/(u(L))^{1/4}$ as our scaling variables~\cite{binder1985finite}. The scaling form of the free energy then depends on $u_0$ which we do not have a way to change or set in the simulation. Instead, we treat $u_0$ as a fit parameter in the scaling form of the susceptibility: 

\begin{equation}
 \chi = L^2 \left(W(y L^{\frac{1}{D}}) + 1\right)^{\frac{1}{2}} \Phi \left( t_0 L^2 \left(\frac{W(y L^{1/D})}{1/(D u_0) - 1}\right)^{-A}\right).
\end{equation}

\noindent At the critical point $t = 0$, the function $\Phi$ must be analytic for
finite $L$ (since non-analyticity requires an infinite system size).
$\Phi(0)$ is therefore a constant independent of $L$ and $u_0$ at $t =
0$. Using this, $u_0$ may be estimated from $\chi$ at different values
of $L$ by fitting to its predicted dependence 
$\chi \propto L^2 (W(y[u_0] L^{1/D}) + 1)^{1/2}$ where 
$y[u_0]$ is defined above.

Figures~\ref{collapses1}--\ref{collapses2} shows the scaling collapse of the magnetization
and susceptibility. The magnetization is collapsed using the best-fit value of $u_0=0.4$. Though our collapses are not significantly
better than the traditional logarithmic forms, the
correct form of the singularity will be more apparent at larger values
of $u_0$. This is because the $\log \log$ term which is the second term in the asymptotic expansion of the $W$ function is very small compared to the $\log$ except at large $u_0$ and small $L$. Changing the value of $u_0$ will require a model different from the nearest neighbor square lattice Ising model.

So far, we have been considering the effects of changing coordinates in the control variables on the predictions of the theory. Wegner~\cite{wegner1974some} had also considered changing coordinates in the degrees of freedom of the theory. These changes lead to `redundant' variables, the corrections from which can be removed by coordinate changes. We discuss them in separate work. Here we merely note that they can be used to explain some features of the scaling, like the fact that the coefficient of the $h u$ term is zero.

\subsubsection{Choice of normal form} 

There are certain choices we have made in our application of normal form theory. One is to keep the flow parameter $\ell$ unchanged. Some of the dynamical systems literature considers changing $\ell$ to depend on other parameters. This would be unusual since the coarse graining length would depend on the physical parameters but does not seem to be disallowed. 
We show in the supplementary material that this does not change the predictions for the 4-d Ising model.

Normal form theory makes a particular choice for what to do with the coefficients that can be changed by coordinate changes: it sets them equal to zero. In general, however, it is not clear that the best choice to make is to set them equal to zero. Consider the equation
\begin{equation}
 \frac{d u}{d \ell} = - u^2 + D u^3 
 \label{ueqchange}
\end{equation}
which, as we saw, has the solution $u(\ell) = 1 /(D (1 + W(y e^{\ell/D})))$. Here, $y = (1/(D u_0)-1) \exp(1/(D u_0)-1)$. Note that $u_0 > 0$ as a requirement for the stability of the free energy. If $u_0 < 1/D$, then $y > 0$, and if $u_0 \leq 1/D$, $y \leq 0$. Hence, the domain of attraction of the fixed point at $u_0 = 0$ has a length $1/D$. If we have a system where $u_0 > 1/D$, then this will lead to $u(\ell) \rightarrow \infty$ in a finite coarse graining length. This is reflected in the branch cut of the $W$ function at $-1/e$. In the context of high energy physics, some have tried to find deep meaning in this pole~\cite{magradze}. 

\begin{figure}
 \includegraphics[width=.85\linewidth]{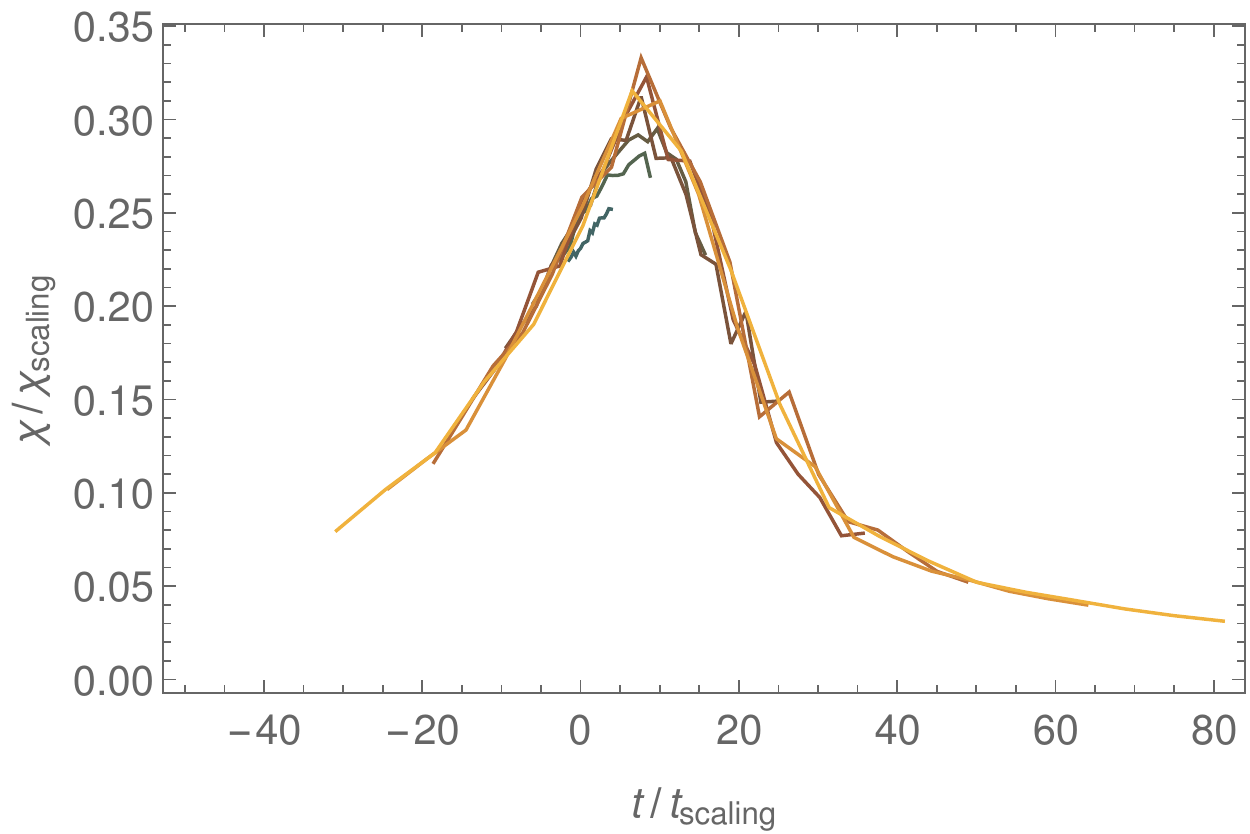}
 \caption{Scaling collapse for the susceptibility using the scaling form given by a different choice of normal form derived from Eq.(~\ref{newnormalform}). Simulations are done on a 4-d lattice using a Wolff algorithm for lattice sizes ranging from $L = 4$ to $L = 32$. Here $\chi_\mathrm{scaling}$ $=$ $L^2 ((W(y L^{1/D}))^{1/2}$ and $t_{\mathrm{scaling}}$ $=$ $L^{-2} (W(y L^{1/D}))^{-1/6}/(1/D u_0))^{1/3} \exp(1/(3 W(y L^{1/D}))) $ with $y = 1/(D u_0) e^{1/(D u_0)}$. We find $u_0 = 0.5$ for the 4-d nearest-neighbor hypercubic-lattice using this method.}
\label{collapses2ndmethod}
 \end{figure}

However, for scaling purposes, we generally prefer a choice of coordinates for which there is no such unphysical behavior. One natural choice is to instead use the equations 
\begin{equation}
 \frac{d u}{d \ell} = - \frac{u^2}{1 + D u} . 
 \label{newnormalform}
\end{equation}
For small $u$, this has the same behavior as Eq.~\ref{ueqchange}. However, the behavior at large $u$ is now well behaved. The solution of this equation is 
\begin{equation}
 u(\ell) = \frac{1}{D W(e^{\ell/D} 1/(D u_0) e^{1/(D u_0)})} ,
\end{equation}
which is in fact somewhat simpler that the solution to Eq.(~\ref{ueqchange}). Scaling collapses with this choice of normal form for the susceptibility are shown in the case of the Ising model in Figure~\ref{collapses2ndmethod}. Better numerics are needed to tell if this choice of normal form is really useful. It turns out that this form has been implicitly used before in the Random Field Ising model as we show explicitly in the next section.

\subsection{Random Field Ising model} \label{sec:randomfieldising}

Finding critical exponents for the random field Ising model has been a longstanding challenge in physics. Some initial results used supersymmetry to prove an equivalence of the Random Field Ising model in dimensions $d+2$ with the Ising model in dimensions $d$~\cite{parisi1979random, de2006random}. It was later shown that the lower critical dimension of the Random Field Ising model is not $3$ (as would be expected from such a correspondence) but rather $2$~\cite{imbrie1984lower}. The upper critical dimension is 6. Here, we will look at the scaling behavior of the Random Field Ising model at its \textit{lower} critical dimension, $d = 2$. 

Consider a spin system with a random field.  
\begin{equation}
 \mathcal{H} = -\sum_{\langle i j \rangle} J s_i s_j + \sum_i h_i s_i ,
\end{equation}
where, $J$ is the nearest neighbor coupling and $h_i$ is a random field chosen from a Gaussian distribution with width $r$. A phenomenological theory for the RG was formulated by Bray and Moore~\cite{Bray85}. It turns out to be useful to define a quantity $w = r/J$. Then, using heuristic arguments on the stability of domain walls, they derive
\begin{equation}
 \frac{d w}{d \ell} = -\epsilon/2 w + A w^3 ,
 \label{rfimwrongeq}
\end{equation}
with $\epsilon = d-2$, and $d$ is the dimension. Note that the flow equations have a symmetry under $w \rightarrow -w$ because the physics is invariant under $r \rightarrow -r$ about the critical point at $r = 0$. This is an example of a pitchfork bifurcation. Bray and Moore argue for this scaling form by looking at the scaling of $r$ and $J$ separately. The scaling of $J$ is given by looking at the energy of a domain wall of size $b^d$. The energy of the domain wall is proportional to $b^{d-1}$. By considering the cost of roughening the domain wall because of the presence of random fields, which goes as $r^2$, they are able to derive the next term in the equation for $J$ which is now
\begin{equation}
 \frac{d J}{d \ell} = (d - 1) J + D w^2 J + \mathcal{O}(w^4) .
 \label{nnequation}
\end{equation}
For the random field $r$, the energy of a region of size $b^d$ is proportional to $b^{d/2}$. Any corrections requires forming a domain of `wrong spins' which, being akin to a barrier crossing problem, is exponentially suppressed. Hence the equation for $r$ is given by 
\begin{equation}
\frac{d r}{d \ell} = \frac{d}{2} r 
\label{rnequation}
\end{equation}
with exponentially small corrections. These two equations together can be used to derive Eq.(~\ref{rfimwrongeq}). Bray and Moore conjecture that Eq.(~\ref{rnequation}( holds exactly to all orders in $w$ (up to exponential corrections). However, it is possible for Eq.~(\ref{nnequation}) to have higher order terms in $w$ and thus Eq.(~\ref{rfimwrongeq}) is only correct to order $w^5$. Integrating Eq.(~\ref{rfimwrongeq}), we get $\ell \sim -1/(2 A w^2) + 1/(2 A w_0^2)$. This implies that the correlation length is 
\begin{equation}
 \xi \sim e^{1/(2 A w_0^2)} .
\end{equation}
For finite size systems, the system size $L \sim \exp(1/(2 A w_0^2)$. Meinke and Middleton~\cite{Meinke05} showed that their finite size data was much better fit by a function of the form $w_0^{-2 y} \exp(C/w_0^2)$ where $C$ is a constant they fit to ($C=\Delta_0$ in their notation) and $y$ = 1.07. We will show that this prediction is consistent with the results of normal form theory. 

As we have already argued, there is no reason Eq.(~\ref{rfimwrongeq}) is true to all orders in $w$. Indeed the, normal form prediction for the flow equations can be derived in a straightforward way. Consider adding a term $A_n w^n$ to Eq.(~\ref{rfimwrongeq}) at $\epsilon = 0$. This is a resonance and cannot be removed usually under normal form theory.  Suppose we make a change of coordinates $w = \tilde{w} + a_n \tilde{w}^{n -2}$. Then, to order $\mathcal{O}(\tilde{w}^n)$, we get
\begin{equation}
 \frac{d \tilde w}{d \ell} = A \tilde{w}^3 + (3 A a_n - A a_n (n -2) + A_n) \tilde{w}^n + \mathcal{O}(\tilde{w}^{n+1}) .
\end{equation}

We can set the coefficient of $\tilde{w}^n = 0$ if we use $a_n = A_n/((n-5) A)$. This procedure fails for $n = 5$ but works for all $n > 5$. \footnote{We note that we are assuming here that the coordinate transformations respect the symmetry of the problem $w \rightarrow -w$. Otherwise, it is possible to remove the $\tilde{w}^5$ term at the cost of introducing a $\tilde{w}^4$ term.} Hence, the normal form of the equilibrium RFIM is given by 
\begin{equation}
 \frac{d \tilde w}{d \ell} = \tilde{w}^3 - D \tilde{w}^5 .
\end{equation}
As before, we have used the freedom to rescale $w$ to set the coefficient of the $\tilde{w}^3$ term to 1. 

The solution of this equation gives us an expression for the correlation length
\begin{equation}
 \xi \sim (1/w^2 - D)^{D/2} e^{1/w^2} .
\end{equation}
This scaling form could explain the data in Meinke and Middleton with $D$ as a fit parameter. Notice that for this to work, $D$ must be positive. However, this solution has the strange property that the correlation length goes to $0$ for $w^2 = 1/D$. If $w^2 > 1/D$, $w^2(l)$ decreases till it reaches $1/D$. If $w^2 < 1/D$, it increases till it reaches $1/D$. As in the 4-d Ising model, it may be more useful to consider instead the flow equation
\begin{equation}
\frac{d \tilde w}{d \ell} = \frac{\tilde{w}^3}{1 + D \tilde{w}^2} .
\end{equation}
This gives the scaling form 
\begin{equation}
 \xi \sim e^{1/(2 w^2)} (w^2)^{-D/2} .
\end{equation}
This is exactly consistent with the scaling form Meinke and Middleton use to collapse their data. Their data would predict the universal value for $D = 2.14$~\footnote{Note that they also have a fit parameter which sets the scale of the exponential. However, this parameter is not universal since it depends on the scale of $w$ unlike $D$}. Any system in the same universality class should see a value of $D$ consistent with this value. However, different values of $D$ would correspond to different universality families within the same class. We now turn to discussing the XY model before returning to the Ising model in dimensions 3, 2, and 1.

\subsection{2-D XY Model}

The 2-d XY model is a remarkable system for several reasons. It was the site
of recently celebrated insight into the connection between ground-state
topology and phase transitions \cite{kosterlitz2017nobel}. Thermodynamic
quantities have essential singularities at its phase transition, not ordinary
power laws, and their derivatives remain continuous to arbitrary order, making
its phase transition infinite order \cite{berezinskii1970destroying,
kosterlitz1973ordering, kosterlitz1974critical}.  This is related the fact
that its RG flow equations are inherently nonlinear: they have no relevant and
two marginal state variables and the procedure laid out by
\eqref{normalformequation} for removing higher order terms from the flow
equations contributes nothing to their simplification.

The XY model is usually posed as ferromagnetically interacting planar spins.
Its partition function is exactly equivalent to the product of a trivial
Gaussian model---corresponding to spin wave degrees of freedom---with a
neutral Coulomb gas---corresponding to the interaction of spin vortices
\cite{knops1977exact}. The latter component contains the interesting critical
behavior, which is characterized by these vortices going through an unbinding
transition. The flow equations for a Coulomb gas in dimension $d$ are given by
\begin{align}
  dK/dl&=-K(\tfrac14Ky^2+d - 2)+\cdots\\
  dy/dl&=-y(K-d)+\cdots,
\end{align}
where $K\sim T^{-1}$ and $y$ is the fugacity of the vortices
\cite{kosterlitz1977d}, which for an XY model is a function of temperature and
cannot be tuned independently but is a free parameter in other equivalent
models, e.g., the Coulomb gas itself. For $d>2$ there is no phase transition
in this system, and for $d<2$ a nontrivial unstable fixed point appears and
there is a phase transition in the hyperbolic universality family. It is worth
noting that these flow equations do not describe the XY model for any
dimension besides $d=2$; 2 is the \emph{upper} critical dimension of the
Coulomb gas and these flow equations, while it is the \emph{lower} critical
dimension for the XY model. At $d=2$ the flow equations undergo a novel
bifurcation: there appears a line of stable fixed points at $y=0$ for all
$K>2$, terminating at $K=2$. This termination is the
Berezinskii--Kosterlitz--Thouless (BKT) critical point. The flow equation near this point with $x=K-2$ is
\begin{align}
  \label{eq:kt-truncated:1}
  dx/dl&=-y^2+\cdots\\
  \label{eq:kt-truncated:2}
  dy/dl&=-xy+\cdots.
\end{align}
These flow equations are zero to linear order and have zero Jacobian at the fixed point.

In principle arbitrary higher-order terms in these equations exist, but there
are several constraints on their form. There is a symmetry $y\to-y$ in the
partition function arising from the neutrality condition---$y$ enters the
partition function in factors of $y^{-\sum_rn_r^2}$ for $\sum_rn_r=0$---which
implies that $dx/dl$ be even in $y$ and $dy/dl$ be odd. In addition, when the
fugacity is zero the model is trivial and $x$ cannot flow, meaning that
$dx/dl$ must only have terms proportional to $y$.  Having applied these
constraints, the simplest normal form has been proven by induction in
polynomial order (Appendix A of \cite{pelissetto2013renormalization}) to take
the form
\begin{align}
  \label{eq:kt-normalform:1}
  d\tilde x/dl&=-\tilde y^2-b_0\tilde x\tilde y^2-b_1\tilde x^3\tilde y^2+\cdots\\
  \label{eq:kt-normalform:2}
       &=-\tilde y^2\big(1+\tilde xf(\tilde x^2)\big)\\
  \label{eq:kt-normalform:3}
  d\tilde y/dl&=-\tilde x\tilde y.
\end{align}
For the BKT point in the sine--Gordon model, which is thought to display to
the same universality as the XY model, it is known that $b_0=3/2$
\cite{balog2000intrinsic, pelissetto2013renormalization}. An infinite number
of coefficients remain, represented here in the form of the Taylor
coefficients of an analytic function $f(x^2)$. These numbers are universal in the
sense that there is no redefinition of $\tilde x$ and $\tilde y$ such that the
flow equations take on the form above and contain different coefficient
values.  Unlike those in the previous sections, this bifurcation does not have
a named classification as far as the authors know. 

A constant of the RG flow can be found by integrating these forms. First,
dividing the equations \eqref{eq:kt-normalform:3} by
\eqref{eq:kt-normalform:2} (and dropping the tildes), we find
\begin{equation}
  \frac{dy}{dl}\bigg/\frac{dx}{dl}=\frac x{y(1+xf(x^2))},
\end{equation}
which separates into
\begin{equation}
  y\frac{dy}{dl}=\frac x{1+xf(x^2)}\frac{dx}{dl}.
\end{equation}
Integrating both sides and choosing $l_0$ such that $x(l_0)=0$, we find
\begin{align}
  \frac12&\big(y(l)^2-y(l_0)^2\big)=\int_{y(l_0)}^{y(l)}y\,dy=\int_{l_0}^ly\frac{dy}{dl}dl\\
  &=\int_{l_0}^l\frac x{1+xf(x^2)}\frac{dx}{dl}dl
  =\int_0^{x(l)}\frac x{1+xf(x^2)}dx.
\end{align}
It follows that 
\begin{align}
  y(l_0)^2&=y(l)^2-2\int_0^{x(l)}\frac x{1+xf(x^2)}dx\\
          &=y(l)^2-x(l)^2+\frac23b_0x(l)^3-\frac12b_0x(l)^4\\
          &\hspace{4em}+\frac25(b_0^3+b_1)x(l)^5+O(x(l)^6)
\end{align}
is a constant of the flow. The expansion of the integral can be taken to
arbitrary order with ordinary computer algebra software. The finite-size
behavior of the flow is rather complicated and doesn't yield closed form
results, details can be found in \cite{pelissetto2013renormalization}.

The XY model and other infinite-order transitions are usually characterized by
the anomalous exponent $\sigma$ parametrizing the essential singularity in the
correlation length,
\begin{equation}
  \xi\sim e^{at^{-\sigma}},
  \label{eq:kt-corr}
\end{equation}
which for the BKT transition is $\sigma=1/2$ \cite{kosterlitz1974critical}.
Conformal field theory predicts the presence of infinitely many models with
this anomalous exponent \cite{ginsparg1988applied}.
The value of $\sigma$ been shown to be fixed by the quadratic-order truncation
of the system's flow equation, independent of any higher-order terms
\cite{itoi1999renormalization}. There are six possible quadratic-order terms in flow equations with two variables. Of these, two can be removed by linear transformations of the two variables. Two more can be set to 1 by rescaling the variables. Hence, there are two parameters at quadratic order which determine the universality family that the system belongs to, and infinite number of subsequent terms which determine the universality class. Giving a full classification of possibilities is beyond the scope of this paper but we give some examples below.

For instance, when the requirement of symmetry under $y\to-y$ is lifted, the
flow equations can no longer be brought to the form Eqs.~\eqref{eq:kt-normalform:2}
and \eqref{eq:kt-normalform:3}; though the simplest form that results isn't
yet known, it is certainly different from the symmetric case, a fact that can
be found by simply trying to eliminate the nonsymmetric cubic terms. In such a
case the codimension of the bifurcation would likely be different, corresponding
to the fact that no \emph{a priori} reason exists for the vanishing of
the term linear in $y$ in $dx/dl$. Such linear terms would change the universality family. Among the infinite collection of BKT-like
conformal theories---including the many physical models identified as having a
BKT-like transition because their behavior resembles Eq.~\eqref{eq:kt-corr}, like
percolation in grown networks \cite{callaway2001randomly,
dorogovtsev2001anomalous}. These may already be examples of models with
$\sigma=1/2$ but belonging to another universality class. It could
also be the case that all BKT-like transitions are in fact members of the same
universality family and class.


Other universality classes and families definitely do exist, characterized by novel values
for $\sigma$. The level-1 $\mathrm{SU}(N)$ Wess--Zumino--Witten model has been
found to be characterized by $\sigma=N/(N+2)$ \cite{itoi1997extended}.
Dislocated-mediated melting alone has produced a melange of anomalous
exponents, with $\sigma=1/2$, $\sigma=2/5$, and $\sigma=0.369\,63\ldots$
depending on precise specification of the model and the lattice geometry
\cite{nelson1979dislocation, young1979melting}. Topological transitions in
systems whose vortices are non-Abelian produce several series of $\sigma$
values dependent on particular symmetry \cite{bulgadaev1999berezinskii}. Each
value of $\sigma$ indicates either a different universality family or merely a different class within the same family depending on how it affects the terms at quadratic order. A classification of possible
bifurcations and corresponding simplest normal forms is in order for flow
equations whose leading order is quadratic, and whose expansions are
constrained or not by various symmetries. This would be the first step in developing techniques for
distinguishing between universality classes and families of this type using
experimental or simulation data. 

\subsection{3-d Ising model} \label{sec:3dising}

There is a sense in which the Ising model is simplest in 3 dimensions because it is part of the hyperbolic universality family. It is also the first natural application of the $\epsilon$ expansion. The transcritical bifurcation at 4 dimensions leads to an exchange of stabilities of the Gaussian fixed point and the Wilson-Fisher fixed point at a non-zero value of $u = u^*$. About this Wilson-Fisher fixed point, the flow equations of the 3-d Ising model are in the hyperbolic universality class with linear coefficients which define the Ising universality class. 

\begin{figure}
 \includegraphics[width=.85\linewidth]{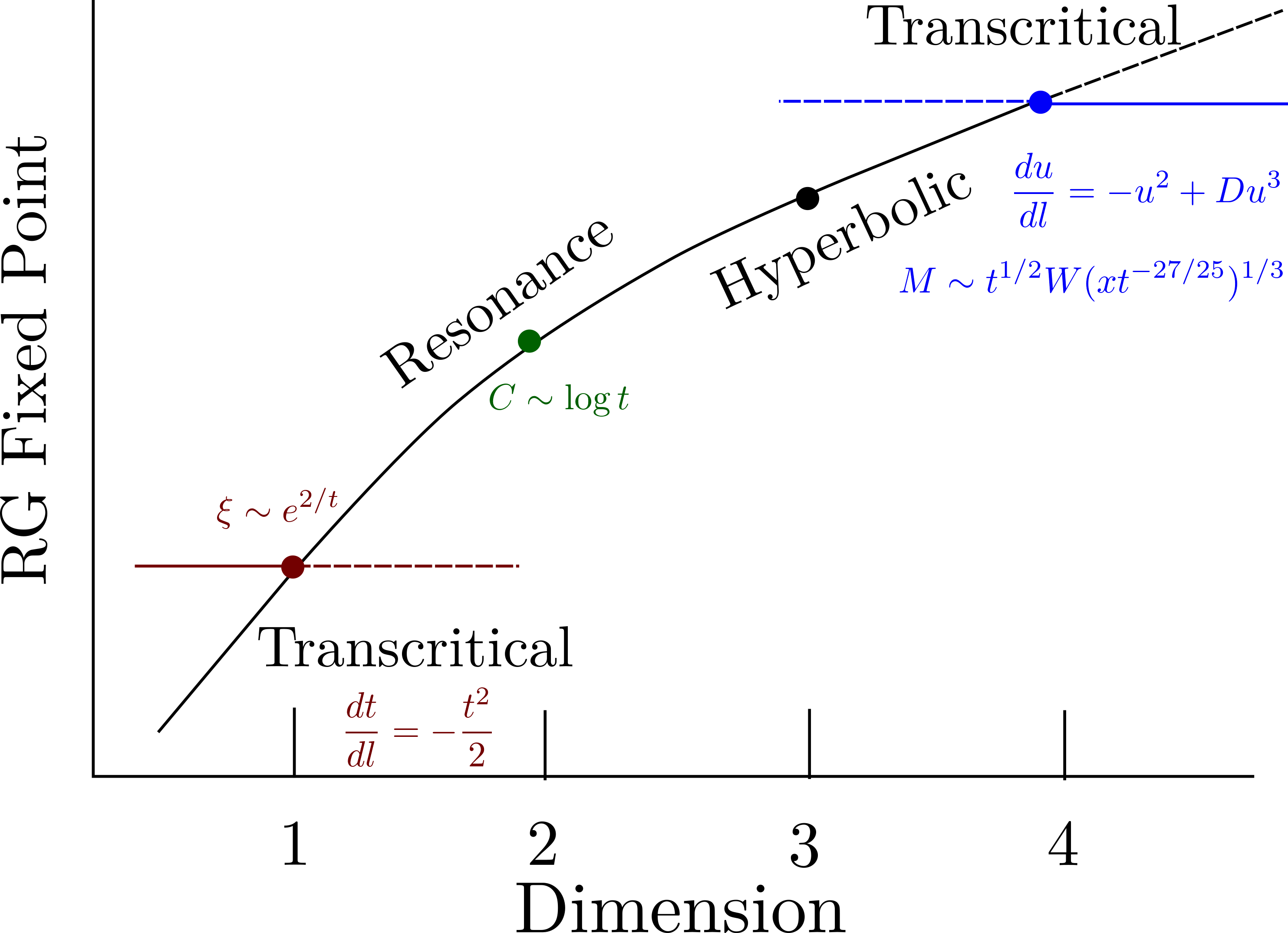}
\caption{Fixed points as a function of dimension in the Ising model. There is a transcritical bifurcation in both 4 and 1 dimensions, leading to $W$ functions and exponential correlation lengths respectively. The fixed point in 3 d is hyperbolic and the flow can be linearized. The fixed point in 2 d has a resonance which leads to a logarithmic specific heat. The challenge is to find a scaling form which interpolates between dimensions giving the correct behavior in all of these dimensions.}
 \end{figure}

However, another approach is to consider the scaling form as a function of the dimension $\epsilon$ in a way that is well defined even at $\epsilon = 0$. Doing this, naturally requires us to keep nonlinear terms in the equation because we already know that the 4-d Ising model has nonlinear terms in its flow equations.

We want to write the flow equations about the 3-d fixed point but keep the nonlinear terms required for the scaling form to have the correct limiting behavior in 2-d and in 4-d. We can write the normal form of the flow equations as 
 \begin{align}
 {d \tilde t}/{d \ell} &= \lambda_t \tilde{t} - A \tilde{u} \tilde{t} ,  \\ 
 {d \tilde u}/{d \ell} &= \lambda_u \tilde{u}- \tilde{u}^2 + D \tilde{u}^3,  \\
 {d \tilde f}/{d \ell} &= d \tilde f - {\tilde t}^2  , \\
 {d \tilde h}/{d \ell} &= \lambda_h h .
\end{align}

We have included the nonlinear terms in $u$ required for the correct scaling behavior and the resonance between the temperature and the free energy. As usual, we switch notation to $t$, $h$ and $u$ with the understanding that they are different from the normal form variables by analytic corrections. Let us look at the scaling variable formed with $t$ and $u$ which can be obtained by solving

\begin{equation}
 \frac{d \tilde t}{d \tilde u} = (\lambda_t \tilde{t} - A \tilde{u} \tilde{t})/(\lambda_u \tilde{u} - \tilde{u}^2 + D \tilde{u}^3) .
 \label{scalingequation3d}
\end{equation}
The solution of this equation gives the scaling variable
\begin{equation}
 t (u)^{-\frac{\lambda_t}{D u_1 u_2}} (u - u_1)^{-\frac{(\lambda_t - A u_1)}{D u_1 (u_1 - u_2)}} (u - u_2)^{\frac{\lambda_u - A u_2}{D (u_2 - u_1) u_2}} = \textrm{const} 
 \label{scaling3d}
\end{equation}
where $u_1$ and $u_2$ are the two non-zero roots of the denominator on the r.h.s of Eq.\ref{scalingequation3d} which to first order in $\lambda_u$ are given by $u_1 = \lambda_u$ and $u_2 = 1/D - \lambda_u$. The form of the scaling variable is interesting, it is essentially given by a product of the linearized scaling variables at the three fixed points that the equation has. Taking the limit $\epsilon \rightarrow 0$, we get 
\begin{align}
 t e^{-2/u} u^{2 D - A} (1 - D u)^{A-2 D} = \textrm{const}
\end{align}
which is the right scaling variable in $4$-d. We have not yet been able to obtain an analytical form for the scaling variable involving $t$ and $h$. This is because the equation for $u(l)$ does not seem to have a closed-form solution here (unlike the 4-d case). Nevertheless, we are motivated by an attempt to create scaling variables which interpolate between different dimensions and have the correct scaling behavior in many dimensions going down from $4$ to $1$. Once the full scaling variables are written down, a first test would be to see if these scaling variables do better collapsing the numerical data in 3-d.

\subsection{1-d Ising model} \label{sec:1dising}

The 1-d Ising model is somewhat different because it is the lower critical dimension and does not have a phase transition. The 1-d Ising model has an exact solution which can be obtained by using transfer matrices. The partition function can be written as the trace of a transfer matrix $\mathcal{T}^N$ where $N$ is the number of spins in the system. The matrix $\mathcal{T}_{i j} = e^{-\beta \mathcal{H}(s_i s_j)}$. Coarse graining here can be done by a well-defined procedure, the coarse grained transfer matrix is defined as $\tilde{\mathcal{T}} = \mathcal{T}^b$ where $b$ is the coarse graining length scale. Defining $\ell = \log b$ and expanding for $b$ close to 1, we can get flow equations for the temperature $T$
\begin{equation}
 \frac{d T}{d \ell} = - \frac{T^2}{2} \sinh \left(\frac{2}{T} \right) \log \left (\tanh \left(\frac{1}{T}\right) \right) .
 \label{1dfullflow}
\end{equation}
This is different from the flow equations we have considered so far because of the presence of non-analytic terms in the flow. The non-analytic term which multiplies the $T^2$ term is $= -1$ at $T = 0$. So, this equation corresponds to a transcritical bifurcation
\begin{equation}
 \frac{d T}{d \ell} =  \frac{T^2}{2} + ...
\end{equation}
where the additional terms are non-analytic at $T = 0$. This can be used to derive a correlation length $\chi \sim \exp(2/T)$.  To interpret the flow further, consider the change of coordinates $\kappa = \exp(-2/T)$. In these variables, the flow is 
\begin{equation}
 \frac{d \kappa}{d \ell} = (\kappa^2 - 1) \log \left( \frac{1-\kappa}{\kappa + 1} \right) .
\end{equation}
Evidently, the flow is analytic in this variable. Solving the full flow Eq.(~\ref{1dfullflow}) gives $\chi \sim -1/(\log \tanh(1/T))$.

For non-zero $\epsilon$, this argument is usually extended in what is called a Migdal-Kadanoff procedure for doing RG~\cite{kadanoff1976notes, chaikin1995principles}. The flow equations are identical except for the presence of a $- \epsilon T$ term which serves as the bifurcation parameter. The $1+\epsilon$ expansion can be summed completely because the flow equation is known to all orders. It does not yield very accurate critical exponents though it gives the exact value of the critical temperature in 2-d (because it respects duality symmetry). Several people have improved the expansion~\cite{martinelli1981systematical, bruce1981droplet}. 

The presence of non-analytic terms in the flow equations complicates the application of normal form theory. We will come back to it when discussing Legendre transform of flow equations.

\subsection{2-d Ising model} \label{sec:2dising}
The 2-d Ising model is a particularly nice example because it has an exact solution in the absence of a magnetic field. All predictions then can be compared to the exact solution. Surprisingly, despite the known exact solution, the scaling behavior of the 2-d Ising model is still not completely understood. A full discussion of the 2-d Ising model will be given in separate work~\cite{Clement18}. Here, we give a brief summary of the issues involved.

The only variable required to describe the 2-d Ising model in the absence of a field is the temperature $t$. The linear eigenvalues of the free energy and the temperature are $2$ and $1$ respectively. The normal form of the flow equations can be written as 
\begin{align}
 \frac{d \tilde f}{d \ell} &= 2  \tilde f - \tilde t^2 , \\
 \frac{d \tilde t}{d \ell} &=  \tilde t .
\end{align}
We have used the fact that the only term which cannot be removed by traditional normal form analysis is the resonance $t^2$. In fact, it cannot be removed by any analytic change of variables. We have also used the freedom to rescale $t$ to set the coefficient of the resonance equal to -1~\footnote{The sign is set to match the exact solution of the square lattice nearest neighbor Ising model}. The solution to this can be written as $\tilde t = \tilde t_0 e^{\ell}$ and the free energy 

\begin{equation}
\label{2disingeq}
 \tilde f(\tilde t_0, \ell) = e^{-2 \ell} \tilde f(\tilde t_0 e^{\ell}) - \tilde t_0^2 \ell .
\end{equation}

Coarse graining until $\tilde t(\ell) = 1$ or $l = -\log(\tilde t_0)$, we get
\begin{equation}
 \tilde f(\tilde t_0) = \tilde t_0^2 \tilde f(1) + \tilde t_0^2 \log \tilde t_0 .
\end{equation}

Now, the normal form variable $\tilde t_0$ is some analytic function of the physical variable $t_0$. It is linear to first order in $t_0$. Hence, we can write it as $\tilde t_0 = t_0 (1 + c(t_0))$ where $c$ is some analytic function. Then, we can expand 
\begin{widetext}
\begin{align}
 \tilde f(t_0) &= t_0^2 (1 + c(t_0))^2 f(1) +  t_0^2 (1 + c(t_0))^2 \log t_0 (1 + c(t_0)) , \\
 &= t_0^2 (1 + c(t_0))^2 f(1) + t_0^2 (1 + c(t_0))^2 \log (1 + c(t_0)) +  t_0^2  (1 + c(t_0))^2 \log t_0 , \\
 &= a(t_0) + b(t_0) \log t_0 
\end{align}
\end{widetext}
where both $a(t_0)$ and $b(t_0)$ are some analytic functions of $t_0$. Meanwhile any change of coordinates which adds an analytic function of $t_0$ to $\tilde f$ can be absorbed in the definition of $a(t_0)$. Hence, we can write the final most general form of the free energy of the 2-d Ising model as $f = a(t_0) + b(t_0) \log t_0$. Indeed, the exact solution of the 2-d Ising model can be written in this form~\cite{caselle2002irrelevant}. 

While the basic solution of the 2-d Ising model is simple, some challenges still remain. The scaling form in the presence of other variables (like the magnetic field and other irrelevant variables) which has so far only been conjectured~\cite{aharony1983nonlinear, caselle2002irrelevant}  naturally follows from an application of normal form theory. It is given simply by including other variables in the argument of the free energy in Eq.(~\ref{2disingeq}) before coarse graining till $t(\ell) = 1$. Irrelevant variables are the source of singular corrections to scaling. An interesting unresolved issue is the presence of higher powers of logarithms in the susceptibility which are not found in the free energy~\cite{orrick2001susceptibility, chan2011ising}. This is usually attributed to the presence of irrelevant variables. Here it is possible to show that the irrelevant variables which are derived from conformal field theory~\cite{caselle2002irrelevant} would in fact lead to higher powers of logarithms in the free energy which are not observed. Hence, they cannot explain the higher powers of logarithms in the susceptibility. It is possible that there are other irrelevant variables in the 2-d square lattice nearest neighbor Ising model with a field which are not predicted by conformal field theory but can capture the higher powers of logarithms in the susceptibility, as they turn on with a field.

The logarithm due to the resonance in the 2-d Ising model is most apparent in the specific heat. It is easy to derive the flow equations for the inverse specific heat which have the form
\begin{equation}
 \frac{d C^{-1}}{d \ell} = 2 C^{-2} ,
\end{equation}
and has a transcritical bifurcation in two dimensions. This raises a question, is it legitimate to talk about a bifurcation in two dimensions for the Ising model if it happens in the space of results rather than the space of control variables? Intriguingly, though perhaps unrelated, a bifurcation has been observed in 2 dimensions using methods of conformal bootstrap~\cite{golden2015no, el2014conformal}. In thermodynamics, a natural framework which interchanges between results and control parameters is given by Legendre transforms. However, the flow equations for the Legendre transformed coordinates generically have non-analyticities in them. We suspect that the variable $t$ (and $h$ etc.) is uniquely specified as the correct variable for RG.  It is possible that it is more natural to consider removing degrees of freedom in the canonical ensemble ($t$ and $f$), then in a microcanonical one ($E$ and $S$)~\footnote{In fact, there is an interesting connection here with information geometry. It is much more natural to talk about the Fisher Information metric in the canonical ensemble. To be able to talk about the uncertainty in a thermodynamic quantity, one has to be able to exchange that quantity with the environment. Otherwise, calculating the Fisher Information Metric can give ill-defined answers. This is further motivation to consider that a particular thermodynamic ensemble may be more suitable for some purposes.}. A fuller discussion will be given in forthcoming work~\cite{Clement18}.

\section{Conclusion}

We have shown how normal form theory leads to systematic procedure for handling the singularity in RG flows. The concept of universality families broadens the notion of a universality class and we have elucidated it with several different examples.  We have focused on getting a precise handle on the singularity at the critical point. However, normal form theory also gives an elegant way to fit corrections to scaling. Interestingly, even the scaling of the 2-d Ising model which has an exact solution has some unresolved mysteries which we are exploring.  It is possible that interpolating between dimensions in a way that
captures the correct singularities can improve scaling collapses
in all dimensions. Finally, we are exploring the application of our methods
to systems like jamming in 2-d~\cite{goodrich2014jamming}, where logarithmic
corrections are observed but no renormalization-group theory is available.  In general, we expect this fruitful confluence of dynamical systems theory and
the renormalization group will not only clarify and illuminate previously
known technical calculations, but will also facilitate quantitative analysis
of experimental and theoretical systems farther from their critical points
and before the underlying field theory is well understood.

\section{Acknowledgements}

We thank Tom Lubensky, Andrea Liu, John Guckenheimer, Randall Kamien and Cameron Duncan for useful conversations. AR, CBC, LXH, JPK, DBL and JPS were supported by the National Science Foundation through Grant No. NSF DMR-1719490. LXH was supported by a fellowship from Cornell University. DZR was supported by the Bethe/KIC Fellowship and the National Science Foundation through Grant No. NSF DMR-1308089.

\bibliographystyle{unsrt}
\bibliography{RGNFbib}

\end{document}